\documentclass[journal,onecolumn]{IEEEtran}

\usepackage{cite}

\title{Towards Advancing Research with Workflows \\
{\large A perspective from the Workflows Community Summit -- Amsterdam, 2025}}

\author{
    \IEEEauthorblockN{
        Irene Bonati\IEEEauthorrefmark{1}, 
        Silvina Caino-Lores \IEEEauthorrefmark{2}, 
        Tainã Coleman\IEEEauthorrefmark{3},
        Sagar Dolas\IEEEauthorrefmark{1},
        Sandro Fiore\IEEEauthorrefmark{4},
        Venkatesh Kannan\IEEEauthorrefmark{5}, \\
        Marco Verdicchio\IEEEauthorrefmark{1},
        Sean R. Wilkinson\IEEEauthorrefmark{6},
        Rafael Ferreira da Silva\IEEEauthorrefmark{6} \\
        \vspace{1em}
    }
    \IEEEauthorblockA{
        \IEEEauthorrefmark{1}SURF Cooperation, Innovation Team, Utrecht, The Netherlands \\
        \IEEEauthorrefmark{2}University of Rennes, Inria, CNRS, IRISA, France \\
        \IEEEauthorrefmark{3}San Diego Computing Center, University of California, San Diego, CA, USA \\
        \IEEEauthorrefmark{4}Department of Information Engineering and Computer Science, University of Trento, Italy \\
        \IEEEauthorrefmark{5}Irish Centre for High-End Computing, Dublin, Ireland \\
        \IEEEauthorrefmark{6}Oak Ridge National Laboratory, Oak Ridge, TN, USA
    }
}
        
\begin{document}

\maketitle

\section*{Executive Summary}

Scientific research increasingly involves multidisciplinary teams, distributed datasets, and computational resources. To address this growing complexity, scientific workflows have emerged as a structured approach to orchestrate computational processes across diverse and distributed resources, handle large datasets and data streams, and ultimately ensure scientific reproducibility and integrity.

The Workflows Community Summit 2025, held in Amsterdam on June 6th, 2025, brought together international experts to explore emerging challenges and opportunities related to scientific workflows, focusing on their impact on infrastructure and knowledge innovation. Identified challenges limiting the adoption of workflows include an imbalance between the generality of workflow systems and their practical utility for multiple diverse scientific applications and domains; the long-term sustainability of workflows as well as workflows systems and services; the lack of recognition and reward systems connected to the development and maintenance of reproducible workflows; and limited standardisation, funding, training as well as collaboration between domain experts, computer scientists and data/computing infrastructure teams. 

Despite these challenges, several action lines can be pursued across multiple levels (technology, policy, and community) to collectively enhance the utility of workflows for scientific discovery. Key areas of focus include:

\begin{itemize}
    \item \textbf{Shifting the perspective for measuring utility, from technology focus to scientific impact:} A fundamental paradigm shift is needed in the computing ecosystem, moving away from a technology-focused design to one that prioritises scientific utility. A growing gap between computing power and its practical utilisation for scientific objectives has become apparent and is not sustainable. Reducing this gap requires a multifaceted strategy incorporating technical development, policy adjustments, knowledge exchange, and community engagement. This shift in perspective will inevitably lead to the definition and adoption of alternative metrics to help assess the influence of application workflows for advancing science. Historically, the focus has been on measuring performance, which assesses raw computational capacity but fails to capture the actual value of computing for science. Instead, we require metrics that better evaluate the real impact of workflows on advancing scientific discovery. 

    \item \textbf{Formalising workflow patterns and benchmarks:} Identifying clear patterns for how scientific workflows should be structured and operated, and applying them across different scientific applications, platforms, and domains, is crucial to ensure they are more transparent, easier to reproduce, and widely usable. Furthermore, to ensure the quality and efficiency of workflows, community-driven benchmarks are needed to evaluate performance, usability, and adaptability.

    \item \textbf{Building a cohesive workflows community:} The adoption of robust scientific workflows benefits all research disciplines. This underscores the importance of building and maintaining a strong international community that covers a range of topics (research, design, orchestration, etc.) related to this theme. Besides serving as a platform for knowledge sharing, establishing such a community can facilitate the definition of shared goals and the harmonisation of efforts (e.g., through the creation of working groups). These efforts should also involve stakeholders from funding bodies and research communities to bring workflows to the attention of scientific and policy agendas.

    \item \textbf{Investing in human capital and expertise:} The growing importance of scientific workflows in advancing discovery underscores the need for greater collaboration and a co-design approach between domain scientists, computer scientists and infrastructure providers. To facilitate this, it is essential to develop workflow engineering roles (and corresponding professional career paths) focusing on the management and governance of scientific workflows throughout their entire lifecycle. Furthermore, to train current researchers and future workflow practitioners, it is recommended to integrate scientific workflow concepts into both computer and computational science curricula,  organise regular and dedicated training for workflow practitioners within sustainable, structured and long-term initiatives at national, continental or international level (i.e., Competence and Training Centres on Workflows for Science and Engineering). 
\end{itemize}

The document begins with a brief introduction outlining the current landscape of the computing ecosystem and the rationale for shifting to a workflow-centric approach. It then highlights the challenges discussed at the summit and concludes by outlining possible action lines.

\section{Current Context}

Computational scientific discovery lies at an inflexion point, driven by rapid digital transformation, fundamental advances in technology and underlying infrastructure, and the human ability to digest, process, and build on existing ideas. Yet, the essence of the scientific process remains unchanged; We observe and synthesise information, design new hypotheses, conduct experiments, perform simulations and analysis, and share results and learnings~\cite{hope2023computational}. 

At the centre of computational scientific discovery is the ability to execute and repeat a variety of scientific tasks using digital tools, data, and advanced computational environments. We refer to this collection of functions executed in a specific pattern or flow as research workflows. As we look towards a future where science increasingly addresses complex questions and adopts trans-disciplinary approaches, a workflow-focused paradigm presents numerous opportunities for excellence and productivity. New developments and technologies present fresh avenues for scientific analysis and experimentation; however, challenges persist in effectively leveraging and orchestrating various tools, existing knowledge, datasets, and results. Workflows, therefore, serve as the bridge between scientific ambitions and the evolving digital infrastructure. 

While science already relies heavily on digital tools, services, and infrastructure, several challenges arise from this approach. Prevailing perspectives often rely on dominant metrics (e.g., FLOPS, BYTES, and GBPS) as the primary indicators of growth and development. While this focus can help meet capacity goals and secure political support, its constant pursuit diverts funding from application engineering initiatives that have driven significant scientific advancements over the past 50 years. Additionally, treating these metrics as primary goals can lead to adverse environmental impacts, which have become increasingly apparent in recent years.

The relentless pursuit of brute-force performance, obtained through patchy performance optimisation, often overlooks the fact that more infrastructure and scale do not necessarily lead to better scientific outcomes. Instead, this can be more expensive, operationally complex, and unsustainable for local communities and biological ecosystems, making it unviable in the long term. 

This notion can be changed by prioritising the design and development of application workflows to serve scientific communities and enable scientific analysis. That's why discussions about next-generation application workflows for research are as crucial as conversations about the development of digital infrastructure. It's therefore equally important to ensure that workflows are resilient and adaptable. This allows them to evolve alongside advancements in computing and data infrastructures, enabling them to exploit emerging compute paradigms such as quantum, neuromorphic, and other unconventional architectures. As a result, the agenda for the next decade must include investments in application design and development. 

Moreover, we possess computational power which, if fully utilised, can be significantly amplified by the proliferation of diverse hardware, complicating workflow optimisation. For example, over the past decade, due to the emergence of diverse use cases and the multidisciplinary nature of science, national HPC centres have increasingly become data-centric and modular. They are transitioning from a monolithic design to one that supports dynamic processing, scales to meet needs, and offers a range of data services within a federated ecosystem. Looking ahead, we must actively pursue co-design between infrastructure solutions, the application ecosystem, and technological advancement, continuously adapting and responding to the evolving needs and challenges of the ecosystem.

\section{Findings of the Workflows Community Summit 2025}

The Workflows Community Summit 2025, co-organised by SURF and the Workflows Community Initiative, supported by ORNL, was held in Amsterdam on June 6th, 2025. The event examined these critical developments and cross-sectoral challenges across five key areas: the convergence of serverless and edge-to-cloud, practices for AI-driven computation, the integration of quantum and classical resources, application-centric HPC infrastructure design, human-centred approaches, and potential policy reform. 

Underpinning these focus areas, the summit addressed essential cross-cutting themes of data management, interoperability, co-design, sustainability, general vs practical utility,  reproducibility and provenance, as well as policy and regulation that impact all workflow paradigms. Through interactive sessions and collaborative discussions, participants explored the future of scientific workflows to serve the evolving needs of researchers while addressing key challenges in scalability, accessibility, and long-term sustainability.

\section{Challenges discussed during the workshop}

The summit highlighted a range of interdisciplinary and cross-sectoral challenges that limit the broader adoption and effectiveness of scientific workflows. 

A first major issue is the lack of \textbf{interoperability} among different initiatives. Many efforts continue to operate in isolation, often using distinct technologies or domain-specific methodologies. This fragmentation results in duplicated work, making it difficult to reuse or build upon existing tools. Additionally, while valuable knowledge and expertise exist, they are often not easily findable. Overviews are typically limited to academic publications and are not well-suited for hands-on use by practitioners. As a result, new tools continue to be created with overlapping functionality rather than extending what is already available. 

A related challenge is the knowledge gap between the domain scientist and the available technology or infrastructure. There is a need for a \textbf{co-design} approach that not only promotes interoperability but also prioritises usability. Emerging tools, such as smart interfaces, natural-language assistants, adaptive workflow builders, and optimisation systems, can simplify complex tasks and lower entry barriers. Such capabilities would enable researchers to focus on the science rather than the technical configuration, and could open the ecosystem to new disciplines and communities.

A second set of challenges relates to the imbalance between \textbf{generality and practical utility}. Abstract and broadly applicable workflow systems can increase accessibility but may fall short in supporting fine-tuned optimisation for specific platforms and tools (and communities). Scientists often require adaptable and efficient workflows that can be tailored to their particular research environments. To address this, there is a growing need for collaboration between domain experts and computer scientists. However, many research institutions lack clear guidance on how to distribute responsibilities between these roles, and the creation of dedicated workflow engineering positions remains limited.

\textbf{Sustainability} is another critical concern. Many current approaches do not support long-term development and maintenance of workflows. Most interdisciplinary teams form around individual projects, often within a single funding cycle, without long-term institutional structures to ensure continuity. This leads to repeated efforts to rebuild or reimplement workflow solutions from scratch. Many researchers still rely on makeshift or fragmented systems, and there is uncertainty about who should be responsible for ensuring quality, documentation, and reproducibility over time. Although we can already see a shift towards more reproducible and FAIR science, researchers are not always incentivised to adopt a workflow-centric approach, especially when working with legacy applications and tools. 

The summit also explored the need to formalise recognition and reward systems for \textbf{reproducible workflows}. Questions arose about the types of data required to support effective workflow sharing and reuse, how to address end-to-end traceability aspects, how to engage researchers more actively in the development and utilisation of workflows, and how to integrate IT-focused workflow engineering expertise into research environments in a more structured and effective manner. 

Finally, \textbf{policy and regulation} issues also present significant challenges, particularly in the context of data sharing and exchange across national jurisdictions. Addressing these obstacles will require coordinated efforts across several technical, institutional, and policy domains. Thus, we need to make changes, but also address the structural causes of problems across social, cultural, and institutional norms, rules, regulations, and paradigms~\cite{o2023fractal}.

\section{Opportunities Ahead for Collective Action}

This section outlines a set of action lines across different levels (technical, policy, and community) that need to be pursued to address the challenges listed above and formalise workflow-driven paradigms for scientific advancements and discovery.

\subsection{Technology harmonisation}

Addressing technological challenges in scientific workflows requires targeted investments in human capital, formalisation efforts, and shared infrastructure. Building expertise around workflows is essential, and it begins with education. Integrating workflow concepts into core curricula of computer and computational science programs will empower a new generation of workflow practitioners to design, deploy, and maintain reproducible computational processes. Embedding these concepts in early training helps normalise workflows as a fundamental element of computational thinking, rather than a peripheral skill. Additionally, it is also essential to provide current researchers and engineers with training opportunities to invest in their skill development. Educational gateways and open-source training platforms~\cite{sefraoui2012openstack, talirz2020materials, mambretti2015next, parashar2023toward} can play a pivotal role in bridging the knowledge gap between users and the complex technologies that underlie modern scientific infrastructure. Developing and leveraging existing resources, such as community-developed learning materials~\cite{suter2026terminology}, can provide a shared language and foundation for teaching and collaboration across disciplines.

Furthermore, defining formal workflow patterns that reflect a wide range of scientific applications is critical. Standardised and machine-assessable models, such as FAIR workflow maturity models~\cite{wilkinson2025applying}, can help evaluate and guide development, improving the transparency, reproducibility, and portability of workflows across platforms and domains.

To ensure the quality and efficiency of workflows, community-driven benchmarks must be established to evaluate performance, usability, and adaptability~\cite{coleman2022wfbench, pauloski2024taps}. These benchmarks provide objective metrics for comparing and improving workflow systems and practices, thereby supporting the broader adoption and continuous improvement of these systems and practices. Promoting open architectures, standard adoption, and portable, reusable software stacks is key to fostering interoperability. The focus on developing specialised components should be supported by a community effort to create cohesive, end-to-end systems that can adapt to evolving scientific and technological needs. By aligning technology development with shared goals and practical challenges, the scientific community can establish a robust and flexible ecosystem to support next-generation research workflows.

Finally, the drive for technology harmonisation should be supported by promoting cross-facility integration and open-access platforms. The Department of Energy National Laboratories in the United States have made significant progress towards cross-facility collaboration through the Integrated Research Infrastructure (IRI) program, and more recently through the Genesis Mission, which aims to significantly increase the productivity and impact of U.S. R\&D by pairing scientists with intelligent systems that reason, simulate, and experiment at extraordinary speed. Achieving this vision depends on robust, reproducible, and portable scientific workflows that connect data, computing, and experimental facilities end to end. Similarly, open-access platform initiatives such as OpenStack~\cite{sefraoui2012openstack}, Materials Cloud~\cite{talirz2020materials}, Chameleon Cloud~\cite{mambretti2015next}, and the National Data Platform~\cite{parashar2023toward} are helping to strategically bridge the infrastructure gap by sharing existing compute and data resources. Providing collaborative and broad access to computational infrastructure not only maximises resource utilisation but also reduces the need for constant, costly upgrades or extensions.

\subsection{Policy innovation}

Shifting policy priorities to support workflow-centric science requires structural investment in application design, development, and long-term maintenance. Dedicated funding calls that prioritise application-level innovation are essential to ensure that scientific workflows receive the same level of support that has historically been reserved for infrastructure. A key recommendation from the summit was to allocate a fixed percentage of infrastructure budgets to application development, thereby strengthening the resilience and effectiveness of the research ecosystem. Furthermore, structural funding at both research group and infrastructure provider levels would enable the hiring of specialised experts and the establishment of sustainable development practices.

Beyond funding, there is an urgent need to foster active co-design between application developers and infrastructure providers. Rather than treating these domains as separate tracks/silos, future policies should support the integrated development models where scientific goals and technological capabilities evolve together. At the same time, it is crucial to broaden the perception of HPC. This technology is not limited to supercomputing or niche academic use cases; it has broad societal relevance that can be further improved when exploited through robust workflows. Policies should communicate this broader value and help create more inclusive narratives around computational science, making space for a diverse set of users, domains, and scientific missions.

\subsection{Knowledge exchange \& community development}

Fostering a vibrant and interconnected workflows community requires continuous knowledge exchange and international collaboration. The creation of international working groups focused on workflows for science can help align fragmented efforts, identify shared goals, and reduce duplication across domains. These groups serve as forums for exchanging best practices, tools, and standards, while also identifying gaps and opportunities for joint development. Building on the success of earlier collaborative infrastructure efforts, the community should now rally around a shared vision for workflows that is equally ambitious but more suited to current scientific needs and digital capabilities.

Making workflows a topic of strategic relevance is essential to embed them within the broader scientific and policy agendas. This requires not only technical coordination but also deliberate engagement with institutional leadership, funding bodies, and research communities. Elevating workflows as a strategic pillar can help shift the focus from short-term tool development to long-term ecosystem building. Launching a movement that places greater emphasis on workflows will therefore drive coordinated actions across disciplines and organisations, increasing the visibility and perceived value of workflow engineering in science.

Community development also involves creating inclusive and accessible spaces for knowledge sharing and exchange. Many workflow practitioners operate without formal communities of practice or clear pathways for recognition and advancement. Establishing dedicated events and platforms (such as the Workflows Community), as well as training opportunities, can lower barriers to entry and accelerate learning for new users. Encouraging interdisciplinary collaboration and building bridges between domain scientists, software engineers, and infrastructure providers will help cultivate a sustainable and diverse community.

Notably, workflows have long played a vital role beyond the scientific domain. In industry, they have been used for decades, powering processes from analytical pipelines to the advanced AI workflows now driving modern production systems. Within research, workflows are also expanding beyond traditional scientific applications, gaining increasing relevance in the humanities. They now support efforts in Indigenous language revitalisation, the restoration and preservation of historical archives and artworks, and other areas where reproducible, data-driven approaches enhance interpretation and conservation. This broadening scope underscores the versatility of workflows and their potential to serve as a unifying framework for knowledge creation and stewardship across disciplines.

 As the field matures, it becomes increasingly important to establish specialised career paths with unique expertise around workflow engineering. The formation of experts in the field will be essential to ensure best practices and support the evolving needs of scientific research. Ultimately, global collaboration on workflows can empower researchers everywhere to pursue more scalable, reproducible, and impactful science.

\section{Future Outlook}

The discussion at the Workflows Community Summit 2025 highlighted that shaping the future of computational workflows will require both technical innovation and sustained community collaboration. To address the identified challenges and capitalise on emerging opportunities, we propose a dual-track action plan. First, we will implement a structured roadmap to advance technology harmonisation, policy innovation, and community development, building on the action lines defined in this memo. This includes establishing technical benchmarks and standards, promoting funding models that allocate a fixed percentage of infrastructure budgets to application development, and creating international working groups to reduce duplication of effort and align global initiatives. Each of these actions will have defined milestones and timelines, ensuring accountability and measurable progress toward a more sustainable and impactful workflow ecosystem.

Looking ahead, the Workflow Community Summit will adopt a regular cycle of two complementary events to maintain momentum and ensure comprehensive coverage of priorities. A virtual summit that will focus primarily on the technical aspects of workflows, targeting the broad workflows community, including developers, researchers, industry and infrastructure providers. This event will serve as a platform for exchanging best practices, presenting new tools, and addressing challenges related to interoperability, scalability, and sustainability. An in-person summit will focus on policy-related topics, including structural funding models, career pathways, educational support, standards adoption, and the integration of workflows into national and international science strategies. The second event, in line with the one summarised in this document, will be designed to engage and influence policymakers, funding agencies, and institutional leaders, translating community priorities into actionable policy recommendations. 

By alternating these events, we can promote continuous community engagement and advance and inform both the policy and the scientific and technological aspects in a coordinated way. This approach will help convert ideas into actionable recommendations, bridge gaps between technical and policy domains, and foster a sustainable ecosystem where workflows are recognised as a key enabler of scientific discovery.

\section*{Acknowledgements}

We would like to thank the co-organisers of the meeting, Ariana Torres Knoop (SURF), Tim Kok (SURF), and Manuel Torres Rodrigues (SURF), for their work and effort in organising and running the event. We also thank all participants of the ``Workflows Community Summit -- Amsterdam, 2025" for their participation and discussions during the meeting, which were essential to the development and content of this paper.

\bibliographystyle{IEEEtran}
\bibliography{references}

\end{document}